\documentclass[lettersize,journal]{IEEEtran}
\usepackage{amsmath,amsfonts,bm}
\usepackage{mathtools}
\usepackage{amsthm}
\usepackage{nicefrac}
\usepackage[colorlinks=true,linkcolor=blue,citecolor=blue,urlcolor=blue]{hyperref} %

\DeclareMathOperator{\E}{\mathbb{E}}

\newcommand{\KL}[2]{\mathrm{KL}(#1 \,\|\, #2)}

\usepackage{xcolor}
\newcommand{\rebuttal}[1]{#1}

\usepackage{amsmath,amsfonts}
\usepackage{algorithmic}
\usepackage{algorithm}
\usepackage{array}
\usepackage[caption=false,font=normalsize,labelfont=sf,textfont=sf]{subfig}
\usepackage{textcomp}
\usepackage{stfloats}
\usepackage{url}
\usepackage{verbatim}
\usepackage{graphicx}
\usepackage{cite}
\begin{document}

\title{Advances in Diffusion-Based Generative Compression}

\author{Yibo Yang, Stephan Mandt
\thanks{YY is with Chan Zuckerberg Biohub. SM is with the University of California Irvine.}%
\thanks{Accepted for publication in \emph{IEEE BITS the Information Theory Magazine}. DOI: \href{https://doi.org/10.1109/MBITS.2026.3725614}{10.1109/MBITS.2026.3725614}. \copyright~2026 IEEE. Personal use of this material is permitted. Permission from IEEE must be obtained for all other uses, in any current or future media, including reprinting/republishing this material for advertising or promotional purposes, creating new collective works, for resale or redistribution to servers or lists, or reuse of any copyrighted component of this work in other works.}
}

\markboth{IEEE BITS the Information Theory Magazine}%
{Yang and Mandt: Advances in Diffusion-Based Generative Compression}

\maketitle

\begin{abstract}
Popularized by their strong image generation performance, diffusion and related methods for generative modeling have found widespread success in visual media applications. In particular, diffusion methods have enabled new approaches to data compression, where realistic reconstructions can be generated at extremely low bit-rates. This article provides a unifying review of recent diffusion-based methods for generative lossy compression, with a focus on image compression. These methods generally encode the source into an embedding and use a diffusion model to iteratively refine it during decoding, so that the reconstruction approximately follows the true data distribution. The embedding can take various forms and is typically transmitted via an auxiliary entropy model, and recent methods also explore the use of diffusion models themselves for information transmission via channel simulation. 
We review representative approaches through the lens of rate-distortion-perception theory, highlighting the role of common randomness and connections to inverse problems, and identify open challenges. 
\end{abstract}


\begin{figure*}[t]
    \centering
\includegraphics[width=0.75\linewidth]{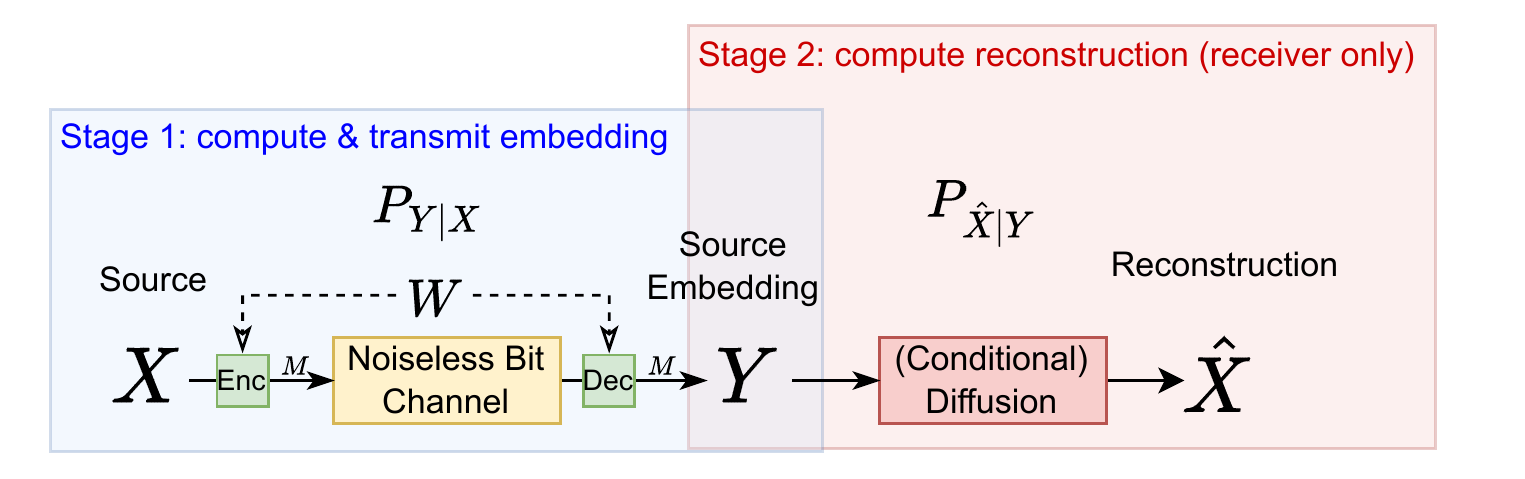}
    \caption{Common architecture of diffusion-based lossy compression methods considered in this article. They follow a two-stage ``compress-then-reconstruct'' approach. In stage 1 (illustrated by the box on the left), an embedding $Y$ of the source $X$ is computed and transmitted from the sender to the receiver. Its behavior can be described by a transition kernel $P_{Y|X}$, which is simulated by methods exploiting a common source of randomness ($W$). For methods based on nonlinear transform coding, this kernel is a deterministic function. In stage 2 (illustrated by the box on the right), the receiver performs additional computation to estimate \rebuttal{a reconstruction} $\hat X$ from $Y$, which can be viewed as solving an inverse problem given a fixed corruption process $P_{Y|X}$. This is commonly done by feeding $Y$ into a (conditional) diffusion/flow-matching model $v_\theta(x_t, y, t)$ and solving a corresponding ODE or SDE. The performance of each method is evaluated based on bit-rate $\E[|M|]$, distortion $\E[\rho(X, \hat X)]$, and divergence (realism, or rather the lack thereof) $\texttt{d}(P_X, P_{\hat X})$. }
    \label{fig:common-architecture}
\end{figure*}

\section{Introduction}

From large language models writing code to video generation tools reshaping digital media, generative AI models are fundamentally changing how we produce and communicate information.
Among these developments, diffusion \cite{sohl2015deep, ho2020denoising, song2020score} and related flow matching methods \cite{lipman2022flow, tong2023improving, albergo2023stochastic} (which we collect under the umbrella term ``diffusion'') have become a dominant approach for modeling and generating high-dimensional data, such as images, videos, audio, and beyond. This success has motivated new approaches to learning-based, or \emph{neural} data compression \cite{yang2023introduction}, where diffusion models can serve as powerful decoders that generate highly realistic reconstructions from compressed representations.
Unlike traditional codecs optimized for pixel-level distortion metrics such as mean squared error (MSE), neural compression methods increasingly aim to achieve high perceptual quality formalized by the notion of \emph{realism} --- the property that the reconstructions are statistically indistinguishable from real data. 
The paradigm shift towards \emph{generative compression}, powered by advances in generative modeling, enables compelling reconstruction quality at extremely low bit-rates \cite{agustsson2019extreme, mentzer2020high, lei2023text+, careil2023towards, theis2022lossy} where conventional approaches produce unacceptable artifacts.

This article provides a unifying review of diffusion-based generative compression methods \cite{yang2023lossy, hoogeboom2023high, lei2023text+, careil2023towards, ghouse2023neural, relic2024lossy, theis2022lossy, yang2025progressive, vonderfecht2025lossy}, with a focus on image compression. Most methods can be viewed as following a two-stage ``compress-then-refine'' architecture (Fig.~\ref{fig:common-architecture}).
In the first stage, the source $X$ is encoded into a compact representation or \emph{embedding} $Y$ and transmitted to the receiver.
In the second stage, a diffusion model is conditioned on this embedding and generates the final reconstruction $\hat X$.
This decomposition holds more generally where the second stage can be any conditional generative model $P_{\hat X|Y}$, and furthermore reveals a close connection between signal \emph{compression} and \emph{restoration} (or \emph{statistical estimation}). Indeed, assuming perfect realism and no common randomness, the optimal second-stage reconstruction under MSE distortion is given by the stochastic inverse of the first-stage encoder (i.e., posterior sampling) \cite{yan2022optimally, hoogeboom2023high}, popularly implemented by learning-based approaches to both \emph{compression} \cite{hoogeboom2023high, theis2022lossy}
and \emph{inverse problems} \cite{daras2024surveydiffusionmodelsinverse}.

A key distinction among different compression methods is whether they exploit \emph{common randomness} --- a shared source of random bits available to both the sender and receiver. 
In one category, methods that do not exploit common randomness can be viewed as instances of nonlinear transform coding \cite{balle2021ntc}. They deterministically encode data into a discrete representation, transmit them via entropy coding, and use a conditional \rebuttal{generative} model for reconstruction.  \rebuttal{Whereas earlier work on generative compression employed GANs as the conditional generative model \cite{agustsson2019extreme,mentzer2020high}, diffusion-based methods in this category \cite{yang2023lossy, hoogeboom2023high, lei2023text+, careil2023towards, ghouse2023neural} essentially upgrade GANs with diffusion models that offer stable training and stronger generation capabilities.}
In the other category, the first-stage encoding $P_{Y|X}$ is stochastic rather than deterministic, and common randomness is exploited by using  \emph{channel simulation} to transmit the representation.
\rebuttal{An unconditional diffusion model is viewed as a hierarchical latent variable model, whose} forward noising process specifies the channel $P_{Y|X}$ to be simulated and its learned reverse process can be used to generate the reconstruction $\hat X$.  This perspective enables lossy compression with a single diffusion model \cite{ho2020denoising} and \rebuttal{uniquely distinguishes methods in this category \cite{theis2022lossy, yang2025progressive, vonderfecht2025lossy} from earlier GAN-based methods \cite{agustsson2019extreme,mentzer2020high}.
Compared to deterministic coding, stochastic coding poses greater conceptual complexity, but bypasses the need for quantization-aware training \cite{agustsson2020universally} and expands the frontier of the best achievable performance limit \cite{theis2021advantages,theis2022lossy}.
By communicating a sequence of decreasingly noisy latents along the diffusion trajectory,} this approach also has an appealing \emph{progressive coding} property, and we show that its communication cost can be elegantly related to the diffusion model's training objective via the celebrated I-MMSE relation of Gaussian channels \cite{guo2005mutual}.

This article is organized as follows. Section~\ref{sec:background} provides background on diffusion modeling, nonlinear transform coding, and channel simulation.
Section~\ref{sec:comm-arch-and-perf-limits} introduces the realism criterion for generative compression, the two-stage architecture, and fundamental performance limits in the tradeoff between rate, distortion, and realism \cite{blau2019rethinking}. Section~\ref{sec:review-of-methods}
reviews representative methods, based on both deterministic coding (no common randomness) and stochastic coding (with common randomness). Section~\ref{sec:conclusions} concludes with a discussion of open problems, regarding improving computational efficiency, generative vs. perceptual compression and evaluation, and bridging the theory and practice of stochastic coding.

Throughout the article, we assume the source (data) is a random vector (r.v.) $X \in \mathbb{R}^n$ following distribution $P_X$. We use the following notation convention. 
We use $\mathcal{N}(\mu, \Sigma)$ to denote a Gaussian measure and its density function with mean $\mu$ and covariance matrix $\Sigma$.
When there is no risk of confusion, we use the shorthand $p(u)$ from machine learning literature to denote the probability density (or distribution) of a r.v. $U$, relying on the lower-case argument $u$ to specify the corresponding random variable. Similarly, for a given conditional distribution (Markov kernel) $P_{U|V}$, we use the shorthand $p(u|v)$ to denote the probability density function (or distribution) associated with $P_{U|V=v}$.

\section{Background}\label{sec:background}

\subsection{Diffusion modeling}\label{sec:background-diffusion}

Diffusion and related methods are now a dominant approach for modeling complex high-dimensional distributions and a key ingredient in recent neural compression methods.
Here we briefly review diffusion models, starting with their original discrete-time formulation as deep hierarchical latent variable models \cite{sohl2015deep}, then viewing them as neural SDEs (stochastic differential equations) in the continuous time limit \cite{song2020score}, and finally extending them to the more general framework of flow matching \cite{lipman2022flow, tong2023improving, albergo2023stochastic} based on dynamic transport of measures.

\paragraph{Diffusion in discrete-time}
A discrete-time diffusion model \cite{sohl2015deep}, commonly known as \emph{DDPM} \cite{ho2020denoising}, is a hierarchical latent variable model that approximates 
the data distribution by inverting a Gaussian noising process.  Starting from the data r.v. $X_0 \equiv X \sim P_X$, the noising process, also known as the \emph{forward process}, is a Markov chain $(X_t)_{t=0,..., T}$ defined by a Gaussian transition kernel $P_{X_t|X_{t-1}=x_{t-1}} =\mathcal{N}(\sqrt{1 - \beta_{t}} x_{t-1}, \beta_{t} I)$, where $(\beta_t)_{t=0, ..., T} > 0$ is an increasing sequence.
It follows that the conditional distribution of $X_t$ given $X_0$ is $p(x_t|x_0) = \mathcal{N}(\alpha_t x_0, \sigma_t^2 I)$, where $\alpha_t$ and $\sigma_t$ are derived from $\beta_{\leq t}$, and $p(x_t|x_0)$ converges to a standard normal distribution as $t \to \infty$.
A diffusion model then learns a variational Markov chain in the reverse direction, called the \emph{reverse process}, parameterized by a fixed \emph{prior} distribution $p_\theta(x_T) = \mathcal{N}(0, I)$ and learned transition kernels $(P^\theta_{X_{t-1} | X_t })_{t=1,...,T}$. As a latent variable model, the model is trained by maximizing the marginal likelihood $p_\theta(x_0)$, which is equivalent to minimizing the expected negative ELBO (Evidence Lower BOund) given on the RHS below,
\begin{equation}
     \mathbb{E}_{x_0 \sim P_X} [ - \log p_\theta(x_0)] \leq \mathbb{E}_{x_0 \sim P_X} [ \mathcal{L}(x_0)], \label{eq:expected-neg-elbo}
\end{equation}
\begin{align}
    \mathcal{L}(x_0): &= \mathbb{E}_{x_1 \sim p(x_{1}|x_0)}[-\log p_\theta(x_0 | x_1) ]  \\ 
    & + \KL{p(x_T | x_0)}{p_\theta(x_T)}  \nonumber \\
    & + \sum_{t=1}^T \mathbb{E}_{x_t \sim p(x_{t}|x_0)} [\KL{p(x_{t-1}|x_t, x_0)}{p_\theta(x_{t-1} | x_t )}]. \nonumber
\end{align}
In the above, the ``forward posterior'' $p(x_{t-1}|x_t, x_0) \propto p(x_{t-1} | x_t) p(x_t | x_0)$ has a closed-form Gaussian distribution, and the reverse-process kernel is chosen to have the same distributional form as $p(x_{t-1}|x_t, x_0)$ but with the value of $x_0$ replaced by a denoised estimate based on $x_t$, i.e., $p_\theta(x_{t-1}|x_t) := p(x_{t-1}|x_t, x_0=\hat x_\theta(x_t, t))$ where $\hat x_\theta(x_t, t)$ is a time-dependent neural network \cite{kingma2023understanding}. 
The KL divergence terms in the objective simplify to squared errors of the form $\mathbb{E} [\|X_0 - \hat x_\theta(X_t, t)\|^2]$, revealing that the optimal predictor is the MMSE estimator $\hat x^*_\theta(x_t, t) = \mathbb{E}[X_0 | X_t = x_t]$.
After training, we can generate samples from the learned data distribution by following the reverse Markov chain.

\paragraph{Continuous-time diffusion modeling: neural SDEs and flow matching} \label{sec:score-sde-fm}
By passing to continuous time, diffusion models can be viewed as neural (stochastic) differential equations that learn to reverse a forward diffusion process \cite{song2020score}. 
In the limit as $T \to \infty$, the noising Markov chain of DDPM becomes a continuous-time Ornstein-Uhlenbeck (OU) diffusion process $(X_t)_{t \in [0, 1]}$ governed by the following SDE:
\begin{equation}
    d X_t = -\frac{1}{2} \beta_t X_t dt + \sqrt{\beta_t} d W_t,
\end{equation}
where $X_0 \equiv X \sim P_X$ and $W_t$ is a Brownian motion. This SDE has a standard solution in terms of the conditional law of $X_t$ given $X_0$,
\begin{equation}
    p(x_t | x_0) = \mathcal{N}(x_t | \alpha_t x_0, \sigma_t^2 I), \label{eq:cond_prob_path}
\end{equation}
where $\sigma_t^2 = 1 - e^{- \int_0^t \beta_s ds }$ and $\alpha_t^2 = 1 - \sigma_t^2$ are collectively known as a \emph{noise schedule}. The marginal law of $X_t$ then has the following density
\begin{equation}
    p_t(x_t) = \int p(x_t | x_0) d P_X(x_0). \label{eq:diffusion_marg_distribution}
\end{equation}
The time-reversal of the forward process can be characterized by another diffusion process, given by the reverse SDE \cite{song2020score}, %
\begin{equation}
    d X_t = [-\frac{1}{2} \beta_t X_t - \beta_t \nabla_{X_t} \log p_t(X_t)] dt + \sqrt{\beta_t} d \bar{W}_t, \label{eq:reverse-score-sde}
\end{equation}
where $\bar{W}_t$ is a Brownian motion running in reverse time, and $\nabla_x \log p_t(x)$ is known as the \emph{score function} at time $t$. 
The key is then to approximate the unknown score function with a time-dependent neural network $s_\theta(x_t, t)$, by solving the following weighted least-squares regression problem,
\begin{equation}
    \int_0^T \lambda(t) \mathbb{E}_{\substack{x_0 \sim P_X, \\ x_t \sim p(x_t | x_0)}} [\|s_\theta(x_t, t) - \nabla_{x_t} \log p(x_t | x_0) \|^2] \, dt, \label{eq:score-sde-denoising-score-matching}
\end{equation}
where $\lambda: [0, T] \to \mathbb{R}^+$ is a positive weighting function. 
After training, we can  generate samples from the learned data distribution by simulating the reverse SDE from time $t=T$ to $t=0$, starting with $X_T \sim \mathcal{N}(0, I)$. %
Alternatively, a deterministic \emph{probability-flow ODE} \cite{song2020score} can be solved to the same effect.
Let $p_\theta(x_0)$ denote the distribution of samples drawn from the learned neural SDE. Then for a suitable choice of $\lambda$, the above \emph{denoising score matching} objective (Eq.~\eqref{eq:score-sde-denoising-score-matching}) can be shown \cite{song2021maximum, kingma2023understanding} to be equivalent to the continuous-time limit of the negative ELBO (Eq.~\eqref{eq:expected-neg-elbo}), which similarly upper bounds the negative likelihood of data under the model $\mathbb{E}_{x_0 \sim P_X} [-\log p_\theta(x_0)]$.

The framework of \emph{flow matching} \cite{lipman2022flow, tong2023improving} (and largely equivalent, \emph{stochastic interpolants} \cite{albergo2023stochastic}) offers a different perspective that subsumes and extends diffusion models. 
Under this perspective, we start by defining a sequence of interpolating variables
$(X_t)_{t \in [0,1]}$
and their associated distributions $(p_t)_{t \in [0,1]}$ exactly as in equations \eqref{eq:cond_prob_path} and \eqref{eq:diffusion_marg_distribution},  but we are now free to design the $\alpha_t$ and $\sigma_t$ functions independently of any forward diffusion process (a common choice being $\alpha_t = 1-t$ and $\sigma_t = t$). This results in a path of probability distributions $(p_t)_{t \in [0,1]}$ that interpolates between the data distribution $p_0 = P_X$ at $t=0$ and a prior Gaussian distribution $p_1=\mathcal{N}(0, I)$ at $t=1$.\footnote{Note that we use the diffusion time convention, which is the opposite in flow matching literature (``noise'' distribution at $t=0$ and ``data'' distribution at $t=1$).} 
Flow matching then trains a time-dependent neural network $v_\theta:  \mathbb{R}^n  \times [0, 1] \to  \mathbb{R}^n$ to approximate a target velocity field known to generate the probability path $(p_t)_{t \in [0,1]}$, by solving the following regression problem
\begin{equation}
    \mathbb{E}_{\substack{t, x_0 \sim P_X, \epsilon \sim \mathcal{N}(0, I), \\ x_t = \alpha_t x_0 + \sigma_t \epsilon}} [\| v_\theta(x_t, t) - (\dot \alpha_t x_0 + \dot \sigma_t \epsilon) \|^2], \label{eq:cfm-objective}
\end{equation}
where $\dot \alpha_t=\frac{d \alpha_t}{dt}$ denotes the time derivative of $\alpha_t$, and similarly for $\dot \sigma_t$.
The learned velocity field $v_\theta$ can then be used to generate approximate samples from the data distribution by solving the following ODE from $t=1$ to $0$, 
\begin{equation}
    d X_t = v_\theta(X_t, t) d t, \quad X_1 \sim p_1.
\end{equation}
In fact, the same procedure works for other choices of prior distributions than the Gaussian, allowing us to learn a neural ODE that bridges between any two continuous probability distributions. 
In the case of a Gaussian prior, flow matching is equivalent to continuous-time diffusion (up to a rescaling of time) with v-prediction and a particular choice of noise schedule \cite{kingma2023understanding}. Moreover, the different prediction targets (denoising prediction, noise prediction, score prediction, and v-prediction) and loss functions in diffusion modeling can all be shown to be equivalent up to the choice of the noise schedule and loss weighting across time \cite{kingma2023understanding}.

\paragraph{Conditional modeling} 
Diffusion modeling as described above can be naturally extended to model a \emph{conditional} distribution $P_{X|Y}$ given paired samples $(X, Y) \sim P_{XY}$. 
To simplify discussion, we adopt the language of flow matching where the central object is a learned time-varying velocity field $v_\theta(x_t, t)$ that %
generates the data distribution,
noting that this is largely equivalent to learning a denoiser $\hat x_\theta(x_t, t)$ or score function $s_\theta(x_t, t)$ as explained above.  
Conceptually, conditional generation then requires a simple modification: instead of learning an unconditional velocity field $v_\theta(x_t, t)$ that generates $P_X$, we now learn a \emph{conditional} velocity field $v_\theta(x_t, y, t)$ that generates $P_{X|Y=y}$ whenever $Y=y$. 
In practice, this involves modifying the neural network $v_\theta(x_t, t)$ to accept an additional input $y$, and the loss function is modified accordingly to train on $(X, Y)$ pairs:
\begin{equation}
    \mathbb{E}_{\substack{t, (x_0, y) \sim P_{XY}, \epsilon \sim \mathcal{N}(0, I), \\ x_t = \alpha_t x_0 + \sigma_t \epsilon}} [\| v_\theta(x_t, y, t) - (\dot \alpha_t x_0 + \dot \sigma_t \epsilon) \|^2]. \label{eq:conditional-cfm-objective}
\end{equation}
An example is learning-based JPEG artifact removal, where $Y$ is the JPEG reconstruction of an image $X$ \cite{saharia2022palette}.

\subsection{Nonlinear transform coding}\label{sec:ntc}
Most existing neural lossy compression algorithms follow the paradigm of \emph{nonlinear transform coding} (NTC), which can be viewed as a data-driven, parametric approach to vector quantization \cite{balle2018hyper, balle2021ntc}. 
The idea is to learn a pair of \emph{analysis} and \emph{synthesis} transforms $(\phi_a, \phi_s)$, as well as an entropy model parameterized by neural networks. The sender maps the input $X$ into transformed coefficients $Y = \phi_a(X)$, and losslessly transmits the quantized coefficients $\lfloor Y \rceil$ under the entropy model, where $\lfloor \cdot \rceil$ denotes coordinate-wise rounding to the nearest integer. The receiver then computes a reconstruction using the synthesis transform $\hat X = \phi_s(\lfloor Y \rceil)$.
To enable end-to-end optimization, the quantization operation is replaced by a uniform noise channel, allowing all components of the system to be optimized on a rate-distortion objective. 
Neural image compression based on NTC has demonstrated better performance than classical codecs on metrics such as PSNR and MS-SSIM \cite{balle2018hyper}, and produces much more pleasing reconstructions especially when augmented with a realism loss (to be discussed in Sec.~\ref{sec:perceptual-realism}).
We refer readers to \cite{balle2021ntc} and \cite{yang2023introduction} for more in-depth discussions of NTC.

\subsection{Channel simulation}\label{sec:channel-sim}
Whereas channel coding essentially turns a noisy channel into a noiseless one, it can be useful to perform the opposite task of \emph{simulating} a noisy channel using a noiseless channel.
Simulating a given noisy channel $P_{Y|X}$ would enable two parties to ``communicate'' through a remote generation scheme, where whenever a sender observes $X=x$, the receiver would observe a sample of $Y$ drawn from the corresponding distribution $P_{Y|X=x}$. 
The technique of \emph{channel simulation} \cite{li2024channel} thus opens up new possibilities for lossy compression beyond the traditional transform coding (and NTC) paradigm, in that it allows describing the source $X$ by a  \emph{stochastic code} $Y|X \sim P_{Y|X}$ that can be easier to design and optimize than a deterministic code based on quantized coefficients \cite[Chapter 3.4]{yang2023introduction}. We give a brief overview of the core concepts and results that are key for information transmission with unconditional diffusion models (Sec.~\ref{sec:uncond-diffusion-methods}).

\paragraph{General setup and basic results}
Given a noiseless binary channel, a conditional distribution $P_{Y|X}$, and a common source of randomness, e.g., an infinite sequence of random bits, a channel simulation algorithm is characterized by the following:
(1) a random variable $W \in \mathcal{W}$ representing the common randomness, available to both the sender and receiver; its distribution  $P_W$ is part of the coding scheme;
(2) an encoding function $f:  \mathcal{W} \times \mathcal{X}  \to \{0, 1\}^* $, where $\{0, 1\}^* := \bigcup_{l=0}^\infty \{0, 1\}^l$; (3) a decoding function $g: \mathcal{W} \times  \{0, 1\}^*   \to \mathcal{Y}$. 
The sender, upon observing $W$ and the input $X \sim P_X$, computes a binary message $M = f(W, X)$ and sends it over the noiseless channel; the receiver receives $M$ and observes the same $W$, and outputs $Y = g(W, M)$. 
A (one-shot) channel simulation algorithm must meet the requirement that $Y$ given $X$ follows the prescribed distribution $P_{Y|X}$, while aiming to minimize the expected description length $\mathbb{E}[|M|]$ and the computational complexity of the procedure. The general setup is illustrated in the left-side box of Fig.~\ref{fig:common-architecture}.
We make the assumption of \emph{unlimited common randomness}, which can be cheaply provided by equipping the sender and receiver with two pseudo-random number generators (PRNGs) initialized to the same seed. Under this assumption, it can be shown that the minimum expected prefix-free code length of one-shot channel simulation, $L^*$, is close to the mutual information $I:=I(X; Y)$ up to a logarithmic factor \cite[Theorem 4]{li2024channel}:
\begin{equation}
    I \leq L^* \leq I + \log_2(I+2) + 3 \text{ bits}, \label{eq:channel-simulation-rate-bound}
\end{equation}
Existing channel simulation algorithms in this setting operate in a way similar to rejection sampling, as follows:
(1). Both the sender and receiver draw an i.i.d. sequence $\bar{Y}_1, \bar{Y}_2, ... \stackrel{\text{i.i.d.}}{\sim} P^\theta_Y $ from a shared reference distribution $P^\theta_Y$ as common randomness. (2). The sender uses its observation $X$ to select an entry $\bar{Y}_K$ from the sequence and transmits the index $K$ to the receiver. (3). The receiver outputs $Y=\bar{Y}_K$, which is guaranteed by the algorithm to follow the target distribution $P_{Y|X=x}$. 
In particular, for any $P_X$, $P_{Y|X}$, and $P^\theta_Y$, \rebuttal{an algorithm known as} \emph{Poisson functional representation} (PFR) \cite[Chapter 3.3]{li2024channel} achieves an expected code length 
\begin{equation}
    L_{PFR} \leq C^\theta + \log_2 (C^\theta+2) + 3 \text{ bits}, \label{eq:pfr-rate-approx-prior}
\end{equation}
where 
\begin{equation}
    C^\theta := \mathbb{E}_{x \sim P_X}[ \KL{P_{Y|X=x}}{P^\theta_Y} ] \label{eq:mi-upper-bound}
\end{equation}
is an upper bound on $I$. By setting $P^\theta_Y$ to be the $Y$-marginal of $P_X P_{Y|X}$,\footnote{In practice this can be nontrivial, and only an approximation to the true $Y$-marginal is obtained by learning a parametric model $P^\theta_Y$ by e.g., maximum-likelihood estimation, hence our notation $P^\theta_Y$.}  the upper bound becomes tight and we recover the RHS of  Eq.~\eqref{eq:channel-simulation-rate-bound}.
For later convenience, we use the shorthand $Y = \operatorname{OCS}(P_{Y|X}, P^\theta_Y, X)$ to denote the output $Y$ of a one-shot channel simulation algorithm for a channel $P_{Y|X}$ given an input r.v. $X$ and a common reference distribution $P^\theta_Y$. %
The logarithmic rate overhead in Eq.~\eqref{eq:pfr-rate-approx-prior} becomes negligible as $C^\theta$ increases, therefore we may want to maximize our communication efficiency by transmitting a large amount of information at a time. 
Unfortunately this is nontrivial: \rebuttal{the expected number of samples required can be shown to satisfy $\mathbb{E}[K] = \Theta(2^{C^\theta})$}
\cite[Chapter 3.5]{li2024channel}, i.e., the computational complexity scales exponentially with the amount of information transmitted. 
One workaround is to split the information into smaller chunks and transmit the chunks independently \cite{flamich2020compressing}. This sacrifices the overall coding efficiency, as each chunk contributes its own logarithmic overhead to the bit-rate; however, it can still achieve strong image compression performance with a latent diffusion model \cite{vonderfecht2025lossy}. 
Instead of allowing an arbitrary channel as input, more efficient channel simulation algorithms can be obtained by making additional assumptions \cite{flamich2023faster}. In particular, an efficient algorithm exists if we restrict ourselves to a uniform noise channel, which we discuss next. %

\paragraph{Dithered Quantization (DQ)}
Consider any $X \in \mathbb{R}^k$ and an additive uniform noise channel $Y = X + U$, where $\Delta > 0$ is a constant and $U \sim \mathcal{U}(-\Delta/2, \Delta/2)^k$ is uniformly distributed on the hypercube $[-\Delta/2, \Delta/2]^k$.
By exploiting the algebraic properties of $\mathbb{R}^k$, \emph{dithered quantization} (DQ) \cite[Chapter 3.6]{li2024channel} allows us to optimally perform one-shot channel simulation in this setting.
In DQ, the common randomness is an independent r.v. $W \sim \mathcal{U}(-\Delta/2, \Delta/2)^k$, with $\Delta$ assumed known to both the sender and receiver. The sender computes a discrete vector $K := \Delta \lfloor \frac{X + W }{\Delta}\rceil $ where $\lfloor \cdot \rceil$ denotes coordinate-wise rounding to the nearest integer, and transmits its binary encoding under the Huffman code for $P_{K|W}$. The receiver decodes $K$ and computes $Y = K - W$. It can be shown that $Y|X$ indeed follows the desired channel distribution, and that the expected coding cost 
satisfies (ignoring the 1-bit overhead of Huffman coding)
\begin{equation}
    \mathbb{E}[|M|] = H(Y|W) = I(X; Y) = h(Y) - k \log_2 \Delta.
\end{equation}
DQ achieves optimal code length, i.e., the mutual information lower bound in Eq.~\eqref{eq:channel-simulation-rate-bound}, and its computation efficiency is agnostic to the amount of information transmitted --- it only requires vector operations and entropy coding. We note that in practical implementations, the $Y$-marginal distribution $P_Y = P_X * \mathcal{U}(-\Delta/2, \Delta/2)^k$ is often not available in closed-form, and is commonly approximated with a density model of the form $P^\theta_Y = P^\theta_X * \mathcal{U}(-\Delta/2, \Delta/2)^k$ \cite{balle2018hyper, agustsson2020universally}; the latter can be estimated by maximum-likelihood, or equivalently minimizing the rate upper bound $C^\theta$ in Eq.~\eqref{eq:mi-upper-bound}. Afterwards, an approximation to the entropy model $P_{K|W=w}$ can be obtained cheaply by discretizing the density of $P^\theta_Y$ on a grid with width $\Delta$ and offset by $w$, ensuring fast entropy coding of $K$. 
When we take $X$ to be the coefficients produced by the analysis transform in NTC (Sec.~\ref{sec:ntc}), DQ gives us a natural alternative to quantization followed by entropy coding, allowing us to 
achieve the same compression performance as implied by the uniform noise channel at training time \cite{agustsson2020universally}.

\section{Generative Compression}\label{sec:generative-compression-framework}
\subsection{Towards perceptual compression with high realism}\label{sec:perceptual-realism}
Many lossy codecs target perceptual signals consumed by humans, e.g., images or audio, so their success arguably depends on the perceptual quality of the reconstructions.
It has long been observed that distortion metrics such as MSE do not capture the perceptual quality of reconstructed images, and there is a large body of work on perceptual metrics; %
we refer to \cite[Sec. 3.5]{yang2023introduction} and \cite{qiu2024wasserstein} for more extensive discussions. 
Compared to classical codecs,  learning-based approaches are more flexible in that they can be end-to-end optimized for custom loss functions, in particular perceptual metrics such as MS-SSIM \cite{balle2021ntc}, and thus can produce reconstructions with significantly higher perceptual quality. 
Here, we consider a popular approach based on the \emph{realism} criterion, which posits that a reconstruction $\hat X$ has good perceptual quality if the distributional divergence $\texttt{d}(P_X, P_{\hat X})$ is small \cite{blau2018tradeoff}. %
Here, $\texttt{d}(\cdot, \cdot)$ is a divergence satisfying $\texttt{d}(\mu, \nu) \geq 0$ for all $(\mu,  \nu)$ and $\texttt{d}(\mu, \nu) = 0$ iff $\mu=\nu$, with a common proxy being FID or KID in image generation/compression.
The rationale is that a realistic-looking $\hat X$ should closely follow the distribution of $P_X$, so that it is impossible for an observer to tell it apart from real data samples. %
The idea has been successfully implemented in learning-based image compression \cite{yang2023introduction} and image restoration \cite{blau2018tradeoff} to achieve high perceptual quality. Commonly this is done by optimizing the decoder with an adversarial loss, such as a GAN loss which approximates the Jensen-Shannon divergence between $P_X$ and $P_{\hat X}$. 
Diffusion-based approaches essentially replace the GAN decoder by a more powerful conditional generative model, and have been shown to achieve higher realism at low bit-rates \cite{yang2023lossy, hoogeboom2023high}. We refer to compression methods operating at (nearly) perfect realism as generative compression methods, which we describe in more detail below.

\subsection{Common architecture and fundamental limits}\label{sec:comm-arch-and-perf-limits}
Most existing generative compression methods can be described by a common architecture following a two-stage procedure, illustrated in Fig.~\ref{fig:common-architecture}.
Given the source r.v. $X$, the first stage extracts an embedding $Y$, which is made available to the receiver via a noiseless binary channel. 
An independent random variable $W$ is optionally available to both the sender and receiver to assist in the transmission of $Y$, as in the setup of channel simulation (see Sec.~\ref{sec:channel-sim}). Specifically, the sender computes a binary message $M \in \{0, 1\}^*$ from $X$ (and optionally $W$), transmits it losslessly, and the receiver decodes $Y$ from $M$ (and optionally $W$). 
In the second stage, the embedding $Y$ is fed as input to a (conditional) diffusion model, which then generates the final reconstruction $\hat X$. %
As we shall see, $Y$ can take many different forms, ranging from a learned quantized tensor \cite{yang2023lossy}, to a text caption \cite{lei2023text+}, or even a pixel-space reconstruction \cite{hoogeboom2023high, ghouse2023neural}. 
Note that in this architecture, the reconstruction $\hat X$ is conditionally independent of the source $X$ given the embedding $Y$, so that the second-stage reconstruction task can be treated from the perspective of solving an \emph{inverse problem} under a given corruption process $P_{Y|X}$. 
Below we consider the performance limit implied by this two-stage architecture, based on the expected bit-rate $\E[|M|]$, distortion $\E[\rho(X, \hat X)]$, and realism (or ``perception index'' \cite{blau2019rethinking}) $\texttt{d}(P_X, P_{\hat X})$, in a three-way \emph{rate-distortion-perception} tradeoff \cite{blau2019rethinking}. 
We divide our discussion based on whether common randomness is used in the communication of $Y$ given $X$. In the first case (deterministic coding), the fundamental performance limit is fairly well-understood in theory, which even suggests a practical construction of an optimal two-stage architecture \cite{freirich2021theory, yan2022optimally, hoogeboom2023high}. The second case (stochastic coding with common randomness) is less well-understood, where a gap exists between theoretically optimal coding schemes and computationally practical algorithms \cite{theis2021coding, theis2022lossy, yang2025progressive}.

\paragraph{No common randomness}
For a given embedding encoder $P_{Y|X}$, the second-stage reconstruction task boils down to estimating $X$ from $Y$ as in an inverse problem, e.g., image restoration. 
It was observed \cite{blau2018tradeoff} that estimators in this setting exhibit a fundamental tradeoff between the achievable distortion and perception (in the sense of realism), as characterized by the
\emph{distortion-perception (D-P) function} $D^{P_{Y|X}}(\gamma)$ \cite{blau2018tradeoff}: %
\begin{align}
    D^{P_{Y|X}}(\gamma) = \inf_{P_{\hat X|Y}: \texttt{d}(P_X, P_{\hat X}) \leq \gamma } \mathbb{E}[\rho(X, \hat X)], 
\end{align}
i.e., for a fixed threshold $\gamma \geq 0$ on the allowable divergence,  $D^{P_{Y|X}}(\gamma)$ yields the best achievable distortion. %
We use the notation $D^{P_{Y|X}}(\cdot)$ to emphasize that the  D-P function here depends on the choice of $P_{Y|X}$.
Although the D-P function possesses a few general properties, %
more insight can be obtained when we focus on MSE distortion and Wasserstein-2 (W-2) divergence. 
In this case, the D-P function admits the following expression \cite{freirich2021theory}
\begin{align}
    D^{P_{Y|X}}(\gamma) = D^{P_{Y|X}}(\infty)+ [ \max(\gamma^* - \gamma, 0)]^2, \label{eq:dp-function}
\end{align}
where $D^{P_{Y|X}}(\infty)$ is the minimum MSE distortion achieved by the unconstrained MMSE estimator $X^* = \mathbb{E}[X|Y]$, and $\gamma^*:= W_2(P_X, P_{X^*})$. It turns out that we can construct a family of optimal estimators $\hat X(\gamma)$ from two estimators operating at two extreme points of the D-P function. 
Let $\hat X_0$ denote any estimator that achieves perfect realism, which can be characterized by the W-2 optimal transport plan between $P_X$ and $P_{X^*}$. Then $\hat X(\gamma)$ defined by the linear combination
\begin{equation}
    \hat X(\gamma): = (1 - \frac{\gamma}{\gamma^*})\hat X_0 + \frac{\gamma}{\gamma^*} X^*, \gamma \leq \gamma^*, \label{eq:d-p-optimal-interpolation-estimator}
\end{equation}
can be shown to traverse the D-P function for each $\gamma \leq \gamma^*$ ($D^{P_{Y|X}}(\gamma) \equiv D^{P_{Y|X}}(\infty)$ is constant and achieved by $\hat X = X^*$ whenever $\gamma \geq \gamma^*$).

Now suppose we fix a bit-rate constraint $R$ and optimize over the choice of encoders $P_{Y|X}$ under the rate constraint. 
This will allow us to obtain the fundamental rate-distortion-perception tradeoff via the D-P function $D^R(\cdot)$ at each rate $R$.
W.L.O.G. we can take $P_{Y|X}$ to be a quantizer that maps $X$ to a quantization index $Y \in \{1, 2, ..., 2^{\lceil R \rceil} \}$, and it can be shown that both the MSE distortion and W-2 divergence are simultaneously minimized when we choose $P_{Y|X}$ to be the MMSE quantizer for the source $P_X$ at rate $R$ \cite{yan2022optimally}. %
In other words, under MSE distortion and W-2 divergence, a rate-distortion optimal encoder is also optimal for the rate-distortion-perception tradeoff at a given bit-rate, and a single MSE-optimized encoder is sufficient for navigating the entire distortion-perception tradeoff on the decoder side. This observation also holds in more general settings and motivates the construction of neural codecs that operate on a \emph{universal representation} \cite{zhang2021universal}. With this choice of encoder, it holds that
\begin{equation}
    W_2^2(P_X, P_{X^*}) = \mathbb{E}[\|X - X^*\|^2] = D^R(\infty), 
\end{equation}
where $D^R(\infty)$ denotes the minimum distortion achieved among all rate-$R$ source quantizers. 
Plugging this into the general expression in Eq.~\eqref{eq:dp-function}, the D-P function becomes
\begin{align}
    D^R(\gamma) = D^R(\infty) + [\max(\sqrt{D^R(\infty)} - \gamma, 0)]^2. \label{eq:rdp-determinsitic-codes}
\end{align}
Thus, the fundamental distortion-perception tradeoff at a fixed rate $R$ can be described in three regimes. 
First, when $\gamma \geq \sqrt{D^R(\infty)}$, the realism (perception) constraint is inactive, and the optimal reconstruction $\hat X$ is the unconstrained MMSE estimator $X^*$.
Second, under the perfect realism constraint $\gamma = 0$, the optimal distortion satisfies $D^R(0) = 2D^R(\infty)$, i.e., it is twice the MMSE distortion at rate $R$ and is achieved by the estimator $\hat X_0$ satisfying $P_{\hat X_0|Y} = P_{X|Y}$, meaning that the decoder performs posterior sampling \cite{yan2022optimally, hoogeboom2023high}. In other words, without access to common randomness, perfect realism requires exactly a doubling of the best achievable MSE distortion at a given bit-rate.
Finally, for any intermediate $\gamma \in (0, \sqrt{D^R(\infty)})$, the optimal reconstruction $\hat X(\gamma)$ can again be given by a linear interpolation between the two estimators $X^*$ and $\hat X_0$ as in Eq.~\eqref{eq:d-p-optimal-interpolation-estimator}.

\paragraph{Exploiting (unlimited) common randomness}
Although generally not necessary for optimal rate-distortion performance, stochastic codes based on channel simulation (see Sec.~\ref{sec:channel-sim}) can strictly outperform deterministic codes when a realism constraint is imposed \cite{theis2021advantages}.
We define the following (informational) \emph{rate-distortion-perception function} \cite{blau2019rethinking}:
\begin{equation}
    R(D, \gamma) := \min_{P_{\hat X | X}: \mathbb{E}[\rho(X, \hat X)]\leq D, \texttt{d}(P_X, P_{\hat X}) \leq \gamma} I(X; \hat X).
\end{equation}
By considering one-shot channel simulation algorithms under the distortion and realism constraints, it can be shown \cite{theis2021coding} that an algorithm exists with an expected code length $L$
\begin{equation}
    R(D, \gamma) \leq L \leq R(D, \gamma) + \log_2(R(D, \gamma)+1) + 4,
\end{equation}
and the logarithmic overhead disappears asymptotically with block coding, thus allowing us to view the rate-distortion-perception function as an ideal operational performance limit.

\subsection{A note on the connection to inverse problems}
We elaborate on the connection between generative compression, as presented in the two-stage approach, and inverse problem solving with diffusion models. 
There are largely two types of approaches for diffusion-based inverse problem solving, both aimed at estimating $X$ from a degraded signal $Y$ by sampling from the posterior $P_{X|Y}$: (1) supervised approaches, which learn a conditional diffusion model of $P_{X|Y}$ based on paired $(X, Y) \sim P_{X} P_{Y|X}$ samples; (2) unsupervised approaches, which do not require paired data and only learn a diffusion prior for $P_X$, then \rebuttal{perform posterior sampling for any given degradation process $P_{Y|X}$ via \emph{inference-time guidance} \cite{daras2024surveydiffusionmodelsinverse}}.
When paired training data is available, supervised approaches often provide better reconstruction performance and computational efficiency compared to unsupervised approaches, but are less flexible and require training a different model for each degradation process.
We refer to \cite{daras2024surveydiffusionmodelsinverse} for a survey of this literature, and note that most diffusion-based compression methods follow the \emph{supervised} approach when solving the stage-two image restoration task. This is because in compression we get to design the ``degradation process'' (i.e., the first-stage encoder $P_{Y|X}$), and paired training data can be easily obtained from it.
However, the test-time adaptability of the unsupervised approach can still be beneficial in enabling novel compression-restoration applications where the performance criteria are only revealed at test time.

\section{Recent Diffusion-Based Generative Compression Methods} \label{sec:review-of-methods}
We now review recent diffusion-based methods for image compression targeting high realism, relating them to the common framework introduced in Sec.~\ref{sec:generative-compression-framework}.
The methods can largely be divided into two categories, based on whether common randomness is exploited. Methods reviewed in Sec.~\ref{sec:cond-diffusion-methods} are based on deterministic coding: they implement the first stage $P_{Y|X}$ by a deterministic function, and the embedding $Y$ (e.g., quantized coefficients produced by an NTC encoder) is made available to the receiver via an entropy-coded discrete representation.
In the second stage, the reconstruction is typically generated from a conditional diffusion model taking $Y$ as the conditioning signal.
Methods reviewed in Sec.~\ref{sec:uncond-diffusion-methods} are based on stochastic coding, which assume and exploit a common source of randomness. %
In such methods, the first stage $P_{Y|X}$ is often the noise corruption (forward) process of an (unconditional) diffusion model of $P_X$, and the embedding $Y$ is a random sample drawn on the receiver side via channel simulation (Sec.~\ref{sec:channel-sim}). The second stage reconstruction is typically generated by following the reverse process of the same (unconditional) diffusion model. As discussed in Sec.~\ref{sec:comm-arch-and-perf-limits}, the choice between deterministic vs. stochastic coding affects the theoretically achievable performance limit, with the latter at an advantage when realism is desired \cite{theis2021advantages}. \rebuttal{We note that it is the use of common randomness, rather than a conditional-vs.-unconditional diffusion architecture, that distinguishes stochastic from deterministic coding. Indeed, not all methods fall cleanly along the architectural divide: one may use quantization and entropy coding instead of channel simulation to transmit a trajectory of noisy latents \cite{ohayon2025compressed} in a deterministic coding scheme similar in spirit to DiffC \cite{theis2022lossy}.
}

\subsection{No common randomness} \label{sec:cond-diffusion-methods}
\paragraph{CDC}
One of the earliest methods, CDC (Yang and Mandt~\cite{yang2023lossy}) can be viewed as NTC (Sec.~\ref{sec:ntc}) with a conditional diffusion model as its synthesis transform. 
In the first stage, as in NTC, the sender computes a tensor of quantized coefficients $Y = \lfloor \phi_a(X) \rceil$ from $X$, %
and then transmits it losslessly under a learned entropy model.
However, in the second stage, instead of applying a one-step synthesis transform, the receiver feeds $Y$ into a conditional denoising network $\hat x_\theta(x_t, y, t)$ and 
samples a reconstruction $\hat X$ from a conditional diffusion model $P_{\hat X|Y}$ aimed at approximating the posterior $P_{X|Y}$.
The model is trained end-to-end on a combination of rate loss and conditional diffusion loss
\begin{equation}
    R^\theta + \E_{t \sim \mathcal{U}(0, T)} [ \lambda(t) \|X_0 - \hat x_\theta (X_t, Y, t) \|^2 ], \label{eq:cdc-loss}
\end{equation}
where $R^\theta$ is the bit-rate estimate of $Y$ under the learned entropy model (see \cite{balle2018hyper} for details), $\lambda: [0, T] \to \mathbb{R}^+$ is a weighting function, $X_0 \equiv X \sim P_X$ and $X_t|X_0 \sim P_{X_t|X_0}$ is defined as in the diffusion forward process (see Sec.~\ref{sec:background-diffusion}).
Although not directly targeting distortion, the conditional diffusion loss effectively acts as an MSE distortion between the source $X$ and intermediate denoised reconstruction $\hat x_\theta (X_t, Y, t)$ aggregated across time, and CDC was shown to achieve superior rate-distortion performance compared to BPG and a strong GAN-based NTC method targeting realism. 
By augmenting the above loss with an additional perceptual distortion loss, where the squared-error function in Eq.~\eqref{eq:cdc-loss} is replaced by a perceptual distortion such as LPIPS, CDC can also be tuned to navigate the distortion-perception tradeoff. %

\paragraph{HFD}
A follow-up approach, HFD (Hoogeboom et al.) \cite{hoogeboom2023high} (and closely related work by Ghouse et al.~\cite{ghouse2023neural}), targets the perfect realism setting and more closely implements the ideal two-stage architecture under MSE distortion and no common randomness. 
As discussed in Sec.~\ref{sec:comm-arch-and-perf-limits}, in this setting a single MSE-optimized encoder in the first stage is theoretically sufficient for the decoder to achieve the full range of optimal D-P tradeoff in the second stage; furthermore, the optimal decoder under perfect realism is given by the posterior sampler $P_{\hat X| Y} = P_{X|Y}$. 
Specifically, in HFD the embedding representation $Y$ is chosen to be the MMSE reconstruction of $X$ for a given bit-rate. This is approximated by an MSE-optimized lossy autoencoder learned using the standard NTC approach (see Sec.~\ref{sec:ntc}). 
The second stage is then conceptually identical to JPEG restoration with a conditional diffusion model \cite{saharia2022palette}, which trains a model of $P_{X | Y}$ on (input, output) pairs from the pretrained first stage model. 
Conceptually, the first stage produces a rate-limited coarse description $Y$ containing high-level structures of the image $X$, whereas the second stage refines $Y$ by generating low-level perceptual details, overcoming the blurriness of MMSE reconstruction observed at low bit-rates. The lack of end-to-end training poses no theoretical limitations to the performance, and the method was shown to outperform the earlier end-to-end approach \cite{yang2023lossy}. This ``compress-then-reconstruct'' approach also has a practical benefit that the MSE reconstruction $Y$ can be relatively cheaply produced as a ``preview'', before the final reconstruction is generated by a more expensive conditional diffusion model. A variant of this approach learns a neural ODE to map from $P_Y$ to $P_X$ with flow matching, which can be seen as an approximation to the $\hat X_0$ estimator in D-P theory \cite{freirich2021theory} and is shown to outperform the conditional diffusion decoder when fewer sampling steps are allowed.

\paragraph{LDM-based}\label{sec:ldm-based-methods}
Other approaches explored the use of pretrained diffusion models, in the form of \emph{latent diffusion models} (LDMs) \rebuttal{\cite{rombach2022high}} where diffusion modeling takes place in the feature space of an autoencoder. 
Text+Sketch \cite{lei2023text+} embeds the input image into a textual description $Y$ %
and implements the reconstruction $P_{\hat X|Y}$ by feeding $Y$ into a pretrained text-to-image LDM. %
It was found that a purely textual representation was ineffective at preserving spatial structures of the input image $X$, so $Y$ is also augmented with a ``sketch'' (edge map) of $X$ when targeting higher bit-rates.
PerCo \cite{careil2023towards} follows a similar approach, where the embedding $Y$ consists of a text caption along with a quantized tensor representation of $X$ extracted from the LDM encoder. The second-stage reconstruction $P_{\hat X|Y}$ is obtained from the LDM similarly to \cite{lei2023text+}, but the diffusion component of the LDM is finetuned on the conditional diffusion loss for $P_{Y|X}$ (see Eq.~\eqref{eq:cdc-loss}) while keeping the LDM autoencoder frozen.
Relic et al. \cite{relic2024lossy} also use a pretrained LDM, but obtain the embedding $Y$ by passing $X$ through the LDM encoder and quantizing the output. 
By varying the quantization bin width $\Delta$, the resulting embeddings $(Y_\Delta)_{\Delta \in [0, \Delta_{max}]}$ are interpreted as corrupted versions of a ``clean'' unquantized embedding $Y_0$ with varying amounts of noise. Thus given $Y_\Delta$, the second stage is viewed as solving an iterative denoising task and performs posterior sampling  $\hat Y | Y_t \sim P_{Y_0|Y_t}$ using the latent-space diffusion model; here $t = \eta(\Delta)$ is predicted by a learned model $\eta$ which approximately maps a given quantization level $\Delta$ to the corresponding time in the forward Gaussian diffusion process. Afterwards, an image-space reconstruction $\hat X$ is obtained by feeding $\hat Y$ through the LDM decoder.
\rebuttal{Relatedly, OSCAR \cite{guo2025oscar} and StableCodec \cite{zhang2025stablecodec} apply nonlinear transform coding in the latent space of an LDM, and propose to speed up decoding by estimating the latent reconstruction $\hat Y$ with the LDM's one-step denoising prediction (``$x_0$ prediction'' \cite{ho2020denoising}) instead of solving a reverse ODE or SDE as usual.}
By exploiting the powerful generation capability of LDMs, these methods have demonstrated increasingly strong image compression performance particularly at low bit-rates ($\leq 0.1$ bits per pixel), exhibiting highly realistic reconstructions.

\subsection{Exploiting common randomness} \label{sec:uncond-diffusion-methods}

The approaches discussed above all face a fundamental performance limit (see earlier discussion under Eq.~\eqref{eq:rdp-determinsitic-codes}): 
Under MSE distortion, their best achievable R-D performance under perfect realism is given by $R(D/2)$ where $R(D)$ is the rate-distortion function of the source, i.e., they necessitate a doubling of the best achievable distortion at a given bit-rate. 
One way to overcome this limit is by augmenting the classical lossy compression setup with a common source of randomness and exploiting it using channel simulation (see Sec.~\ref{sec:channel-sim}). 

\paragraph{Lossy compression with Gaussian diffusion}\label{sec:diffc}
The idea of transmitting information with a single unconditional diffusion model was introduced in DDPM \cite{ho2020denoising}, and later refined in DiffC \cite{theis2022lossy}.
As we shall see, an (unconditional) diffusion model of the source $P_X$ has a natural connection to lossy compression\footnote{In the case of \emph{lossless} compression, diffusion models can also be understood as optimizing the coding cost of lossless compression equal to the negative ELBO \cite{kingma2021variational}. We discuss this more in the context of Algorithm~\ref{alg:diffc}.}
that we make mathematically precise below. 

\rebuttal{\emph{Theoretical considerations.}} Recall that starting from the source r.v. $X_0 \equiv X \sim P_X$, a continuous-time diffusion model defines a corruption process $(X_t)_{t \in [0, T]}$ so that at each time $t$
\begin{equation}
    X_t = \alpha_t X_0 + \sigma_t N_t,  \quad t \in [0, T],
\end{equation}
where $N_t \sim \mathcal{N}(0, I)$ is an independent Gaussian noise. 
For this Gaussian channel, we can connect mutual information with mean-square estimation error through the fundamental I-MMSE relation \cite{guo2005mutual},
i.e., it holds that
\begin{equation}
    \frac{d}{d \xi} I(X_0; X_t) = \frac{1}{2} \operatorname{mmse}(\xi). \label{eq:immse}
\end{equation}
where $\xi:= (\frac{\alpha_t}{\sigma_t})^2$ is the ``SNR'' (signal-to-noise ratio) of the channel, and the RHS of the equation is the minimum MSE
\begin{equation}
    \operatorname{mmse}(\xi):=\min_{\hat x_0: \mathbb{R}^n \times \mathbb{R}^+ \to \mathbb{R}^n} \mathbb{E}[\|X_0 - \hat x_0(X_t, \xi)\|^2 ],
\end{equation}
achieved by the conditional expectation $\hat x_0(x_t, \xi) = \mathbb{E}[X_0 | X_t = x_t]$. 
In diffusion and flow matching, $\xi$ is a strictly monotonically decreasing function of $t$, so that $\xi \to 0$ as $t \to T$, and $X_T$ approaches an isotropic Gaussian noise containing no signal of $X_0$.
Now fix a $\tau \in (0, T]$, and consider varying $t \in [\tau, T]$ so that $\xi$ varies in the range $[\xi_T, \xi_\tau]$. 
Integrating the I-MMSE relation (Eq.~\eqref{eq:immse}) across $\xi$ yields
\begin{equation}
    I(X_0; X_\tau) = \underbrace{I(X_0; X_T)}_{\approx 0} + \int_{\xi_T}^{\xi_\tau} \frac{1}{2} \operatorname{mmse}(\xi) d\xi. 
\end{equation}
We have now written $I(X_0; X_\tau)$ precisely in terms of the canonical time-averaged denoising MSE objective of diffusion (and flow matching with a Gaussian prior) \cite{kingma2023understanding}.
The high-level idea of lossy compression with Gaussian diffusion is then to view $X_\tau = \alpha_\tau X_0 + \sigma_\tau N_\tau$ as a lossy representation of the source $X_0$, and transmit $X_\tau$ given $X_0$ by simulating the Gaussian channel $P_{X_\tau | X_0}$ with the help of a pretrained diffusion model. Afterwards, a reconstruction $\hat X$ can be estimated on the basis of $X_\tau$. In view of the two-stage framework (see Sec.~\ref{sec:comm-arch-and-perf-limits}), this corresponds to the second stage $P_{\hat X|Y}$ with source ``embedding'' given by $Y=X_\tau$. %
Note that the mutual information $I(X_0; X_\tau)$ is the irreducible communication cost (see Sec.~\ref{sec:channel-sim}), so training a continuous-time diffusion model can be viewed as optimizing an upper bound on this cost.

We start by considering the theoretical merit of this approach and examine if it is worth implementing at all. Theis et al. \cite{theis2022lossy} gave an analysis of the rate-distortion performance in the perfect realism setting ($P_{\hat X} = P_X$), 
assuming an ideal, asymptotic implementation of channel simulation so that the bit-rate is $I(X_0; X_\tau)$.
Given $X_\tau$ simulated from $P_{X_\tau | X_0}$, two types of reconstructions $\hat X$ are considered, both corresponding to posterior sampling $\hat X | X_\tau \sim P_{X_0 | X_\tau}$ under the assumption of perfect modeling\footnote{I.e., we assume that we learn a perfect time-reversal of the forward diffusion process, so that our model distribution $P^\theta_{X_0 | X_\tau}$ equals the true posterior $P_{X_0 | X_\tau}$.}: 1). stochastic reconstruction $\hat X_{\text{SDE}}$ based on solving the reverse-process SDE (Eq.~\eqref{eq:reverse-score-sde}), and 2). deterministic reconstruction $\hat X_{\text{ODE}}$ based on solving the probability-flow ODE \cite{song2020score}. It was shown that with an appropriately modified diffusion process, for a multivariate Gaussian source the R-D performance of $\hat X_{\text{SDE}}$ is given by $R(D/2)$, i.e., the same as the best possible performance that can be achieved without common randomness (as discussed earlier in Sec.~\ref{sec:comm-arch-and-perf-limits}), and this holds in an approximate sense for an approximately Gaussian source. Moreover, for any continuous source with a sufficiently smooth density, the MSE distortion of $\hat X_{\text{ODE}}$ is only half of that of $\hat X_{\text{SDE}}$ in the high-rate limit, corresponding to a 3 dB better SNR \cite{theis2022lossy}. Together, these results suggest that $R(D)$ may be achievable at high bit-rates whereas approaches based on \rebuttal{deterministic coding} (reviewed in Sec.~\ref{sec:cond-diffusion-methods}) can only achieve $R(D/2)$ at best \cite{theis2022lossy}.

\rebuttal{\emph{Algorithmic aspects.}} Despite the theoretical potential of the approach, implementing it in practice has been nontrivial. Below we focus on the practical and algorithmic aspects, where it is common to work with diffusion in discrete time through the lens of a latent variable model (see Sec.~\ref{sec:background-diffusion}). Let $\delta > 0$ be the step size of time discretization, and for convenience we take $T$ and $\tau$ to be multiples of $\delta$. As before, $T$ stands for the end time of the forward diffusion process and $\tau$ is a fixed time step where we want to transmit $X_\tau$.
Applying a channel simulation algorithm to simulate $P_{X_\tau | X_0}$ typically requires a common reference distribution that approximates the $X_\tau$-marginal. 
Unfortunately, $P_{X_\tau}$ is a highly complex mixture distribution (see Eq.~\eqref{eq:diffusion_marg_distribution}) and a diffusion model does not provide us with an explicit density for it. 
However, a diffusion model does give us a model of $X_{t - \delta} | X_t$ in the form of a Gaussian density $p_\theta(x_{t-\delta}| x_t)$, which in turn gives us an explicit joint density model $P^\theta_{X_{\tau: T}}$,
\begin{equation}
    p_\theta(x_\tau, x_{\tau+\delta}, ..., x_T) = p_\theta(x_T) \prod_{t={\tau+\delta, \tau+2\delta, ...T}} p_\theta(x_{t-\delta} | x_t). \label{eq:diffc_joint_prior}
\end{equation}
One idea is then to apply channel simulation to the channel $P_{X_{\tau: T}|X_0}$ (which has a joint Gaussian distribution), using $P^\theta_{X_{\tau: T}}$ as the reference distribution, which indirectly accomplishes our goal of simulating $X_\tau|X_0 \sim P_{X_\tau | X_0}$.
This achieves a coding cost upper bounded by $C^\theta + \log_2(C^\theta+2) + O(1)$ (as in Eq.~\eqref{eq:pfr-rate-approx-prior}), with $C^\theta$ given by
\begin{align}
    C^\theta &= \mathbb{E}_{x_0 \sim P_X}[\KL{P_{X_{\tau: T}|X_0=x_0}}{P^\theta_{X_{\tau: T}}}] \\
    &= \sum_{t={\tau+\delta, ..., T}} C_t^\theta, 
    \label{eq:diffc_rate_upper_bound_as_nelbo}
\end{align}
where we define
\begin{equation}
    C_t^\theta := \begin{cases}
        \mathbb{E}[\mathrm{KL}(p(x_T | x_0) \| p_\theta(x_T))], \hspace{-4pt} & t = T, \\
        \mathbb{E}[\mathrm{KL}(p(x_{t-\delta}|x_t, x_0) \| p_\theta(x_{t-\delta} | x_t))], \hspace{-4pt} & \tau\!+\!\delta \leq t < T,
    \end{cases}
\label{eq:diffc-per-step-kl-cost-def}
\end{equation}
and both expectations integrate over $x_0 \sim P_X$, and the second one also over $x_t \sim p(x_t | x_0)$.
Note that Eq.~\eqref{eq:diffc_rate_upper_bound_as_nelbo} equals the negative ELBO loss of DDPM (Eq.~\eqref{eq:expected-neg-elbo}) without the log-likelihood term involving $p_\theta(x_0| x_\delta)$. %
While this idea works in principle, a practical implementation faces at least two severe computational challenges.
First, it is computationally expensive to evaluate or draw a single sample from the reference distribution $P^\theta_{X_{\tau: T}}$, required by one-shot channel simulation. 
Evaluating or sampling from each density model $p_\theta(x_{t-\delta}| x_t)$ already requires an expensive neural-network call, and doing so repeatedly for $P^\theta_{X_{\tau: T}}$ costs as much as generating from the diffusion model up to time $\tau$.
Second, known algorithms for exact one-shot simulation of multivariate Gaussian channels require drawing $\Theta(2^{C^\theta})$ many samples (see Sec.~\ref{sec:channel-sim}), so that the computational cost of the overall procedure would be hopelessly high. 
We can address the first challenge by taking a divide-and-conquer approach, breaking the joint variable $X_{\tau: T}$ into its smaller, constituent chunks $X_\tau, X_{\tau + \delta}, ..., X_T$. Specifically, instead of simulating $X_{\tau: T}|X_0$ all at once, we first simulate $X_T$ from $p(x_T|x_0)$ under the reference distribution $p_\theta(x_T)$, and then iteratively simulate from $p(x_{t-\delta}|x_t, x_0)$ under the \emph{conditional} reference distribution $p_\theta(x_{t-\delta} | x_t )$ conditioned on the  $X_{t}$ already transmitted in the previous step; this is repeated for $t = T-\delta, T-2\delta, ...$ until $X_\tau$ is simulated at $t=\tau + \delta$. 
The receiver can then generate a reconstruction $\hat X$ on the basis of $X_\tau$.
We formalize this procedure in Algorithm~\ref{alg:diffc} (recall the $\operatorname{OCS}$ shorthand from Sec.~\ref{sec:channel-sim}).
For clarity of presentation, we abstract away the $\operatorname{OCS}$ operation as a black box, but each of these operations accepts a source of randomness (e.g., a synchronized PRNG) as an additional input and emits a binary message over the noiseless binary channel.
We require the $\operatorname{OCS}$ algorithm to use prefix-free codes, so that the binary message from each $\operatorname{OCS}$ call can be transmitted and decoded by the receiver as it becomes available. Thus we see that the algorithm naturally implements \emph{progressive coding} --- instead of waiting to receive the entire bit-stream, the receiver at any time $t$ can use $X_t$ and the bits received so far to produce an incremental reconstruction (``preview''), which becomes more informative of the original data $X_0$ as more bits arrive.
A small modification to the algorithm further allows it to support lossless compression when the source $X_0$ is discrete: set $\tau = \delta$, and entropy-code $X_0$ given the already simulated $X_\delta$ under an entropy model  $P^\theta_{X_0|X_\delta}$. This step contributes an additional coding cost of $\mathbb{E}_{(x_0, x_\delta) \sim P_X P_{X_\delta|X_0}}[-\log p_\theta(x_0|x_\delta)]$, resulting in an overall lossless coding cost equal to the negative ELBO objective of DDPM (Eq.~\eqref{eq:expected-neg-elbo}).\footnote{We note that there is a different approach for lossless compression with an unconditional diffusion model which also targets the negative ELBO coding cost, based on \emph{bits-back coding} instead of channel simulation \cite{kingma2021variational}.}
From a practical standpoint, the progressive coding property is very attractive as it allows a codec to flexibly adapt to changing bandwidths and greatly enhances the user experience, especially in real-time applications such as video streaming. 
Compared to methods based on NTC (Sec.~\ref{sec:ntc}), the unconditional diffusion-based approach also simplifies codec deployment: variable-rate compression can be achieved by training and deploying a single diffusion model, instead of one set of neural-network-based encoder, decoder, and entropy models for each target bit-rate.

\begin{algorithm}[t]
\caption{Progressive lossy compression with an unconditional diffusion model and a one-shot channel simulation algorithm ($\operatorname{OCS}$). \textbf{Inputs}: source $X$, time discretization step size $\delta>0$,  end-of-transmission time $\tau \in \{\delta, 2\delta, ..., T\}$. \textbf{Outputs}: reconstruction $\hat{X}$. }\label{alg:diffc}
\begin{algorithmic}[1]
    \STATE Let $X_0 \equiv X$
    \STATE \emph{\# Transmit} $X_T | X_0 \sim P_{X_T|X_0}$
    \STATE Let $X_T = \operatorname{OCS}(P_{X_T|X_0}, P^\theta_{X_T}, X_0)$
    \FOR{$t = T, T-\delta, \dots, \tau+\delta$}
        \STATE \emph{\# Transmit} $X_{t-\delta} | X_t, X_0 \sim P_{X_{t-\delta}|X_t, X_0}$
        \STATE Let $X_{t-\delta} = \operatorname{OCS}(P_{X_{t-\delta}|X_t, X_0}, P^\theta_{X_{t-\delta}|X_t}, (X_t, X_0))$
    \ENDFOR
    \STATE Estimate a reconstruction $\hat X$ based on $X_\tau$, e.g., $\hat X \sim P_{X|X_\tau}$
    \STATE Return $\hat{X}$
\end{algorithmic}
\end{algorithm}

Despite its conceptual appeal, the diffusion-based progressive coding algorithm (Algorithm~\ref{alg:diffc}) still faces a few practical hurdles. By breaking the simulation of $X_{\tau: T}$ into a sequence of smaller channel simulation tasks of $X_{t-\delta}$ for each $t$, our coding cost has become worse, with each of the $C_t^\theta$ summand in Eq.~\eqref{eq:diffc_rate_upper_bound_as_nelbo} contributing its own logarithmic overhead (see discussion in Sec.~\ref{sec:channel-sim}). With small $C_t^\theta$ and many time steps (common implementations of DDPM use $\sim 1000$ time steps \cite{ho2020denoising}), this coding overhead adds up. 
On the other hand, at certain time steps $C_t^\theta$ can still easily be too large  (e.g., over a hundred or thousand bits per step near $t=0$) for the corresponding $\operatorname{OCS}$ operation to be feasible, given its fundamentally $\Theta(2^{C_t^\theta})$ sampling complexity.
Recent work \cite{vonderfecht2025lossy} introduced practical workarounds to these issues. The high-level idea is to keep the amount of information transmitted in each $\operatorname{OCS}$ operation in a ``sweet spot'' that balances coding efficiency and computational efficiency (in this work, the sweet spot is determined to be around 16 bits).
To reduce the logarithmic coding overhead when $C_t^\theta$ is very small (especially near $t =T$), we skip the channel simulation of $X_{t-\delta}$ at these time steps. E.g., suppose $T=1, \delta=0.01, \tau=0.1$; instead of simulating $X_{\tau: T} = (X_{0.1}, X_{0.11}...,X_{0.98}, X_{0.99}, X_1)$, we may choose to only simulate $(X_{0.1}, X_{0.11}, ..., X_{0.9}, X_1)$ with a coarser time discretization near $t=T$, assuming that the Gaussian reverse-process model $p_\theta(x_s | x_t)$ still provides a good approximation to $P_{X_s|X_t}$ for $(s, t)$ that are farther apart.
To reduce the computation cost when $C_t^\theta$ is too high, we can split up the (usually high-dimensional) $X_{t-\delta}$ vector into smaller chunks and perform one-shot channel simulation on them independently \cite{flamich2020compressing}. 
When applied to the latent space of a pretrained LDM \cite{rombach2022high}, this approach demonstrated competitive image compression performance at very low bit-rates against existing generative compression baselines.

\paragraph{Progressive coding with uniform channel simulation}
A different approach, UQDM (Yang et al.~\cite{yang2025progressive}), introduced a modified diffusion process that bypasses the fundamental difficulty of Gaussian simulation entirely, resulting in an efficient implementation of the progressive coding Algorithm~\ref{alg:diffc} based on dithered quantization (DQ). 
Recall that the main difficulty of implementing Algorithm~\ref{alg:diffc} with Gaussian diffusion stems from the repeated  simulation of the Gaussian channel $P_{X_{t-\delta}|X_t, X_0}$. Whereas existing algorithms for exact one-shot simulation of multivariate Gaussian channels have exponential running time in the amount of information transmitted, uniform noise channels can be very efficiently simulated exactly using DQ (as reviewed in Sec.~\ref{sec:channel-sim}). 

Yang et al. \cite{yang2025progressive} therefore replace the Gaussian channels with uniform noise channels, modifying the standard diffusion formalism both in the forward (noising) process and the reverse process to ensure an efficient implementation of the $\operatorname{OCS}$ operation with DQ.  %
Unlike in a Gaussian diffusion model whose forward process is typically specified by the transition kernel $P_{X_{t+\delta}|X_t}$ (or, alternatively by the perturbation kernel $P_{X_t|X_0}$), in UQDM we specify a forward process by the collection of $X_0$-conditioned channel distributions that appear in Algorithm~\ref{alg:diffc}, consisting of $P_{X_T|X_0}$ and $(P_{X_{t-\delta}|X_t, X_0})_{t=\delta, 2\delta, ..., T}$.
In particular, UQDM keeps the Gaussian $P_{X_T|X_0=x_0} = \mathcal{N}(\alpha_T x_0, \sigma_T^2 I)  \approx \mathcal{N}(0, I)$, but chooses a uniform distribution for each $P_{X_{t-\delta}|X_t, X_0}$ with the same mean and variance as in the Gaussian case.\footnote{Interestingly, with these choices it can be shown that the resulting $P_{X_t|X_T}$ for any $t$ has the same mean and variance as in Gaussian diffusion, and approaches the corresponding Gaussian distribution in the continuous time limit \cite{yang2025progressive}.} E.g., the width of $P_{X_{t-\delta}|X_t, X_0}$ is given by $\Delta_t = \sqrt{12 \varsigma_t^2}$ where $\varsigma_t^2$ is the variance of $P_{X_{t-\delta}|X_t, X_0}$ in Gaussian diffusion and is determined by the noise schedule.
As in Algorithm~\ref{alg:diffc}, the reverse process in UQDM provides reference distributions $P^\theta_{X_T}$ and $(P^\theta_{X_{t-\delta}|X_t})_{t=\delta, 2\delta, ..., T}$ used by the channel simulation steps. Here, $P^\theta_{X_T} = \mathcal{N}(0, I)$ as in a Gaussian diffusion model. The choice of the reverse-process kernel $P^\theta_{X_{t-\delta}|X_t}$ requires more care. As discussed in Sec.~\ref{sec:channel-sim}, a computationally efficient way to implement DQ is to choose its reference distribution $p_\theta(x_{t-\delta}|x_t)$ to have the form $p_\theta(x_{t-\delta}|x_t) = \psi_\theta(x_{t-\delta}|x_t) * \mathcal{U}(-\Delta_t/2, \Delta_t/2)^n$, where $\psi_\theta(x_{t-\delta}|x_t)$ is an underlying density model parameterized similarly to the reverse-process kernel of a Gaussian diffusion model (e.g., via denoising prediction or noise prediction), and $\Delta_t$ is the width of the uniform noise channel $P_{X_{t-\delta}|X_t, X_0}$ in the forward process. 

With these choices, we can now implement the progressive coding Algorithm~\ref{alg:diffc} efficiently. Only two small adjustments to the algorithm need to be made compared to the Gaussian diffusion case. First, we skip the initial $\operatorname{OCS}$ call on line 3; instead we use common randomness to draw the same $X_T$ sample from the reference distribution $P^\theta_{X_T} = \mathcal{N}(0, I)$ on both the sender and receiver side. This step has 0 communication cost and can be seen as simulating a degenerate channel $P_{X_T|X_0}$ that equals the reference distribution $P^\theta_{X_T}$, justified by ensuring the SNR $\xi(t)= (\frac{\alpha_t}{\sigma_t})^2$ of the forward process is sufficiently close to 0 at $t=T$. Second, we implement the subsequent $\operatorname{OCS}$ calls (on line 6) with DQ, instead of a general channel simulation algorithm like PFR \cite[Chapter 3.3]{li2024channel}. Crucially, this step has constant (instead of exponential) computation complexity in the amount of information transmitted (see discussion in Sec.~\ref{sec:channel-sim}), since 
DQ is not based on a sampling scheme like PFR. 
To obtain a progressive reconstruction $\hat X$ (on line 8), UQDM proposes to solve the same reverse-process SDE or ODE in discrete time as in the Gaussian diffusion approach \cite{theis2022lossy}.
The resulting coding cost of Algorithm~\ref{alg:diffc} is exactly equal to $C^\theta$ given by Eq.~\eqref{eq:diffc_rate_upper_bound_as_nelbo}, with $C_T^\theta = 0$ and each DQ step contributing the corresponding $C_t^\theta$ (Eq.~\eqref{eq:diffc-per-step-kl-cost-def}). Note that unlike with Gaussian simulation, there is no longer a per-step logarithmic rate overhead. 
As a result, we can end-to-end optimize the coding cost by training the reverse-process UQDM model directly on Eq.~\eqref{eq:diffc_rate_upper_bound_as_nelbo}, via standard stochastic gradient descent. 
As discussed earlier, we can also support lossless compression of discrete $X_0$ by optimizing the model on the full negative ELBO (Eq.~\eqref{eq:expected-neg-elbo}) and entropy coding $X_0$ at the end of Algorithm~\ref{alg:diffc}.
One difficulty with the UQDM approach is that the empirically optimized coding cost (as given by the negative ELBO) does not monotonically improve with an increasing number of time discretization steps, whereas the opposite holds for Gaussian diffusion models \cite{kingma2021variational}. This likely has to do with the inability of the chosen form of reverse-process model in UQDM to well approximate the forward process in the continuous time limit. Nonetheless, UQDM demonstrated competitive rate-distortion and rate-realism performance against neural and classical baselines on progressive image compression. %

\section{Conclusions} \label{sec:conclusions}

In this article, we reviewed essential ingredients for neural lossy compression with diffusion models, including different (but largely equivalent) formulations of diffusion modeling, nonlinear transform coding, channel simulation, and rate-distortion-perception theory, and discussed representative diffusion-based lossy compression methods in the neural image compression literature. 
These methods can be viewed through a common two-stage architecture (Fig.~\ref{fig:common-architecture}), whose performance can be mathematically understood through the rate-distortion-perception tradeoff \cite{blau2019rethinking}.
The first stage encodes the source input $X$ into an informative embedding $Y$, and communicates it to the receiver via either entropy coding or channel simulation. %
The second stage then estimates a reconstruction $\hat X$ from the embedding $Y$ purely on the receiver side, by feeding the embedding into a (conditional) diffusion model. 
For a fixed first stage, the second stage essentially corresponds to solving an inverse problem whose performance is characterized by a fundamental distortion-perception tradeoff \cite{blau2018tradeoff}, and the availability of common randomness in the first stage in turn lifts the overall best achievable rate-distortion-perception performance \cite{theis2021coding, theis2022lossy}. 

Most existing methods operate without common randomness and are based on entropy coding a discrete representation, following the nonlinear transform coding paradigm \cite{balle2021ntc}. Under MSE distortion and perfect realism, the optimal second-stage reconstruction is achieved by posterior sampling from $P_{X|Y}$ \cite{yan2022optimally}, which is popularly implemented by training a conditional diffusion model on $(X,Y)$ pairs \cite{yang2023lossy, hoogeboom2023high, ghouse2023neural, careil2023towards}.
\rebuttal{Here, deterministic coding faces a fundamental performance limit of $R(D/2)$, whereas stochastic coding with common randomness can do strictly better \cite{theis2021coding, theis2022lossy}.
This motivates lossy compression with a single diffusion model of $P_X$ exemplified by DiffC \cite{theis2022lossy} and UQDM \cite{yang2025progressive}. In these methods, the representation} $Y$ is a noise-corrupted version of $X$ produced by a noising (forward) process, and $P_{\hat X|Y}$ is implemented by solving a neural (stochastic) differential equation in the reverse process.
The key is to exploit common randomness and transmit $Y$ by simulating the noisy channel $P_{Y|X}$ as defined by the forward process. We show that the fundamental communication cost of this procedure, $I(X; Y)$, can be exactly related to the loss function of the diffusion model of $X$.
Although the exact simulation of high-dimensional Gaussian channels remains an open problem, practical workarounds \cite{vonderfecht2025lossy} and uniform-noise diffusion formalism exploiting dithered quantization \cite{yang2025progressive} have recently \rebuttal{enabled practical lossy compression with}
a single diffusion model. The resulting progressive coding algorithm offers strong rate-distortion-perception performance controllable by a single bit-stream and the potential to drastically simplify the development and deployment of neural codecs.

Despite their promising performance, diffusion-based generative compression methods face many open problems that drive future research. We outline some key topics below:
\paragraph{Computational complexity}
A major limitation of diffusion-based compression is the high computational cost of decoding, which requires iteratively sampling from a (conditional) diffusion model by solving an ODE or SDE over many time steps.
While the encoding procedure is relatively cheap (e.g., a single forward pass through an analysis network in NTC-based methods, or noise corruptions in DiffC-based approaches), the iterative decoding procedure can require hundreds of neural network evaluations, making real-time or resource-constrained applications challenging.
Several directions have been explored to address this issue. First, recent advances in few-step and one-step generation (see e.g., \cite{boffi2025flow} and references therein) are making it possible to generate high-quality samples with a few (or even single) ODE/SDE steps. These techniques can be readily adapted to diffusion-based compression to drastically speed up the generation of a stochastic reconstruction. 
Second, in methods based on deterministic coding where an R-D optimal representation is also sufficient for optimal rate-distortion-realism (see Sec.~\ref{sec:cond-diffusion-methods} and  Sec.~\ref{sec:comm-arch-and-perf-limits}), the R-D optimal reconstruction in the first stage can be produced cheaply using an asymmetric NTC architecture \cite{yang2020improving, yang2023computationally} or even a classical codec, offering a fast ``preview'' that can be subsequently refined by a more expensive diffusion reconstruction in the second stage for improved realism.
Third, by applying diffusion modeling in the latent space of an autoencoder, latent diffusion offers another way to reduce the computational cost of diffusion-based reconstruction (see Sec.~\ref{sec:cond-diffusion-methods}). Despite demonstrating strong realism at ultra-low bit-rates \cite{lei2023text+, careil2023towards}, this approach comes with additional engineering complexity (by requiring a separate autoencoder) and a fundamental
performance limitation: the best achievable reconstruction quality is upper-bounded by that of the autoencoder itself, which may be suboptimal for compression at higher bit-rates.

\paragraph{Generative reconstruction fidelity and evaluation}
With increasingly large and powerful generative models and rapid methodological improvements in diffusion research, the ideal of generative compression achieving perfect realism becomes increasingly within reach. 
However, this also introduces the dual challenge of  controlling the reconstruction fidelity and properly evaluating the performance of generative compression methods. 
Generative reconstructions can appear perfectly realistic yet differ substantially from the original input data, especially at low bit-rates. 
E.g., diffusion-based reconstructions can suffer from a loss of mid- to high-frequency image details when conditioned on an MSE-based first-stage reconstruction \cite{hoogeboom2023high}, and entire large-scale spatial structures are hallucinated at extremely low bit-rates \cite{lei2023text+, careil2023towards}.
This behavior was also exhibited in earlier GAN-based methods \cite{agustsson2019extreme, mentzer2020high} and is not limited to diffusion-based methods.
To ensure the generated reconstruction remains perceptually faithful to the original image, one can condition the generative decoder on additional information such as edge/color maps \cite{lei2023text+, careil2023towards}, semantic label maps \cite{agustsson2019extreme}, or saliency/ROI (Region of Interest) maps predicted using a saliency model \cite{balle2025good}.
In the high realism and low-rate regime, traditional distortion metrics such as PSNR and MS-SSIM (and even learning-based metrics such as LPIPS) become ineffective indicators of generative reconstruction quality \cite{careil2023towards}. Indeed, generative compression methods targeting extremely low bit-rates (e.g., $\leq 0.01$ bits per pixel) produce highly realistic reconstructions yet significantly underperform conventional codecs such as BPG or JPEG on the PSNR metric \cite{lei2023text+, careil2023towards}.
While human assessment remains the gold standard of perceptual quality, it can be expensive to scale. A few surrogate metrics have emerged for measuring the preservation of semantic and spatial information, such as CLIP score and IoU (Intersection over Union) metrics for object detection and segmentation tasks \cite{careil2023towards}. Rooted in models of human visual perception, another approach termed \emph{Wasserstein distortion} \cite{qiu2024wasserstein} aims to unify and generalize distortion and realism, and has served effectively as an optimization objective for neural image compression \cite{balle2025good}.
Developing scalable (and ideally, optimizable) metrics of reconstruction quality that are aligned with human perception remains an important direction.

\paragraph{Improved stochastic coding}
By augmenting the classical setup of lossy compression with an additional common source of randomness, stochastic codes can perform no worse than deterministic codes, and can perform strictly better when realism is desired. Although this better performance has been demonstrated theoretically in a few examples \cite{theis2021advantages, theis2022lossy}, to what extent this holds generally remains unknown. Moreover, the theory says little about practical, algorithmic implementations. As a result, the design of neural compression algorithms that best exploit common randomness %
remains largely an art, compared to deterministic coding where the optimal architecture is known in some common cases \cite{yan2022optimally, freirich2021theory, hoogeboom2023high}.
Bridging theory and practice would require the development of computationally efficient channel simulation algorithms \cite{flamich2023faster}, as well as improved neural compression algorithms built on top of them \cite{vonderfecht2025lossy, yang2025progressive}. Given the fundamental computational complexity of simulating general channels \cite{li2024channel, agustsson2020universally}, 
future neural compression algorithms can benefit from principled applications of \emph{approximate} channel simulation (see e.g., \cite[Chapter 3.4, Chapter 8]{li2024channel} and references therein) or leverage channels that are efficient to simulate, e.g., by looking beyond the Gaussian channel \cite{agustsson2020universally, yang2025progressive}.

\section*{Acknowledgments}
Stephan Mandt acknowledges funding from the National Science Foundation (NSF) through an NSF CAREER Award IIS-2047418, IIS2007719, the NSF LEAP Center, the Chan Zuckerberg Initiative, and the Hasso Plattner Research Center at UCI.

\bibliographystyle{IEEEtran}
\bibliography{references}

\begin{IEEEbiography}[{\includegraphics[width=1in,height=1.25in,clip,keepaspectratio]{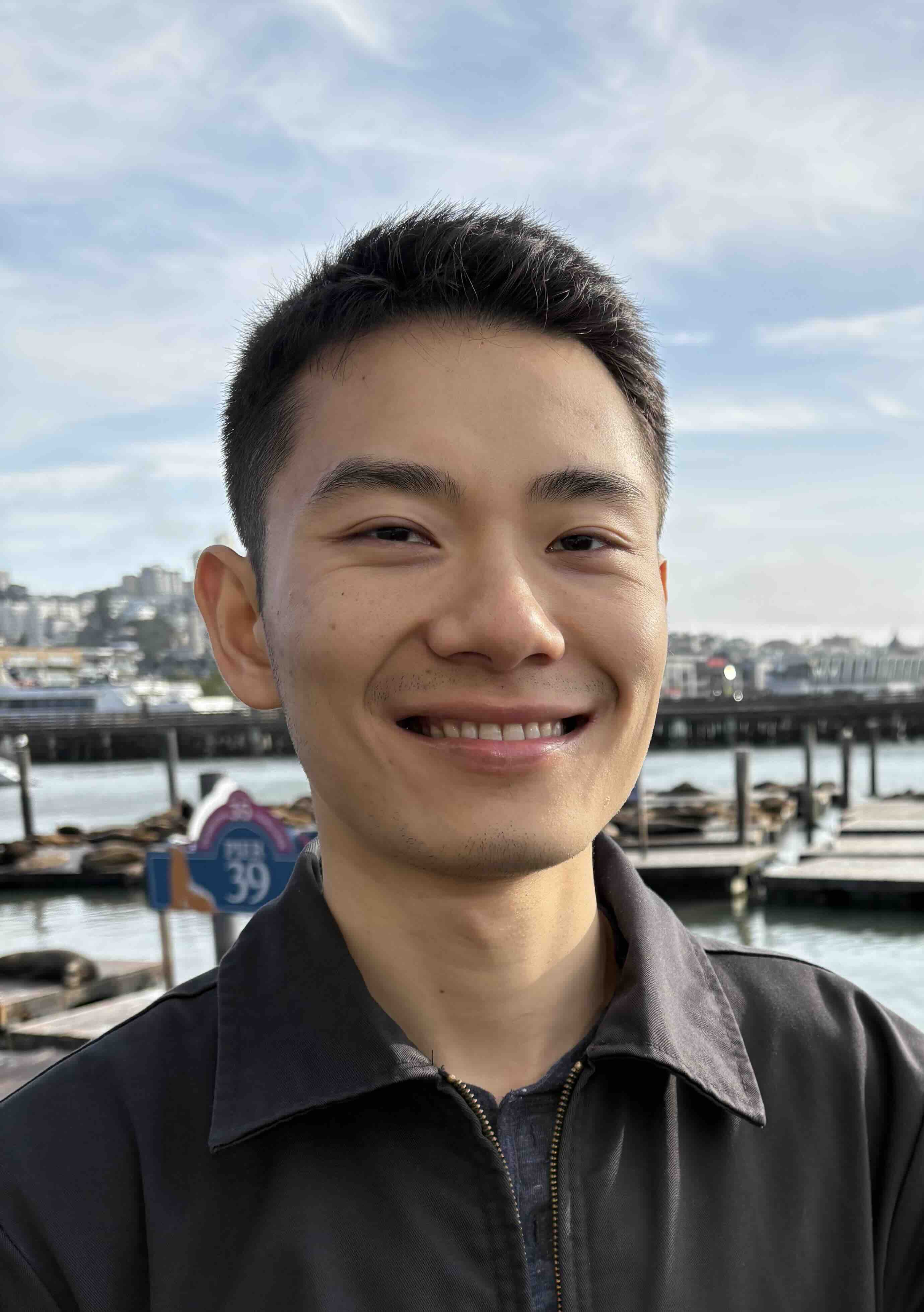}}]{Yibo Yang} is a research scientist at Chan Zuckerberg Biohub working on AI and machine learning for scientific discovery. He received his Ph.D. in Computer Science from the University of California, Irvine in 2024, advised by Stephan Mandt. His doctoral work established fundamental connections between neural (learned) compression, deep generative modeling, and information theory, and he has led numerous tutorials and workshops in this area.
\end{IEEEbiography}

\vspace{11pt}

\begin{IEEEbiography}[{\includegraphics[width=1in,height=1.25in,clip,keepaspectratio]{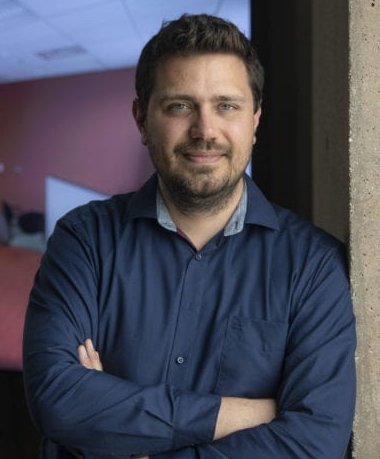}}]{Stephan Mandt} is a Professor of Computer Science and Statistics at the University of California, Irvine. His research contributes to the foundations and applications of generative AI, with a focus on generative modeling of 2D, 3D, and sequential data, compression, resource-efficient learning, inference algorithms, and AI-driven scientific discovery. He is a Chan Zuckerberg Investigator and AI Resident and has received the NSF CAREER Award, the UCI ICS Mid-Career Excellence in Research Award, and a Kavli Fellowship. Before UCI, he led the machine learning group at Disney Research and held postdoctoral positions at Princeton and Columbia. Stephan frequently serves as a Senior Area Chair for NeurIPS, ICML, and ICLR and was most recently Program Chair for AISTATS 2024 and General Chair for AISTATS 2025.
\end{IEEEbiography}

\vfill

\end{document}